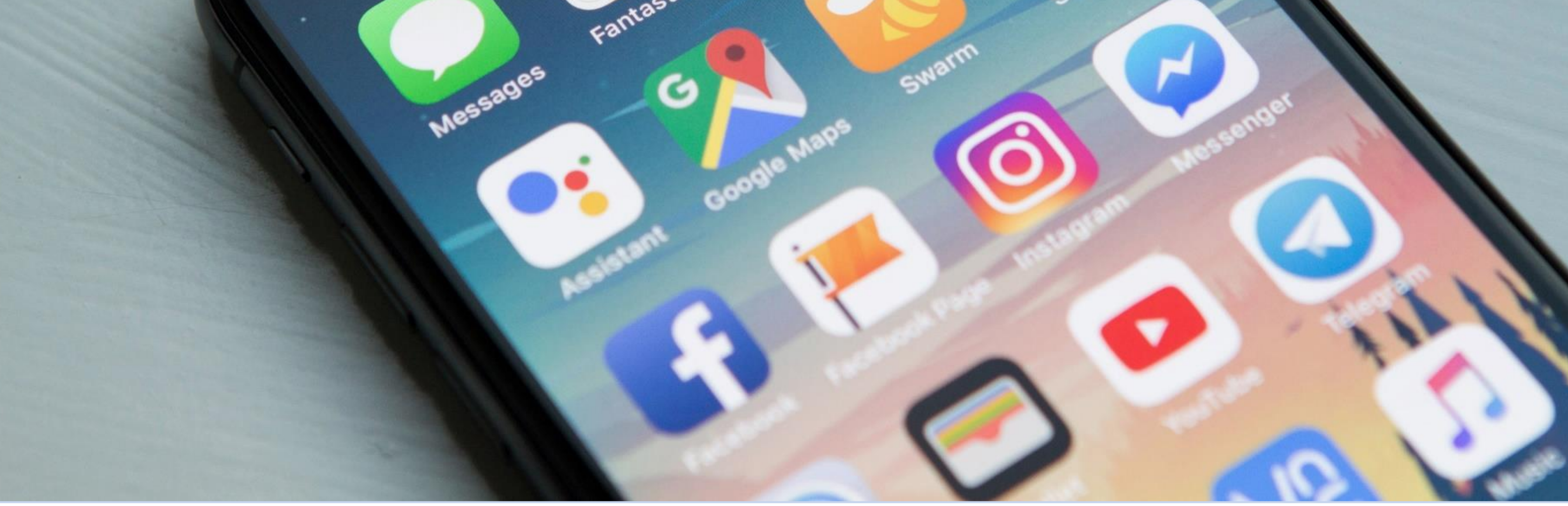

# The social media use of adult New Zealanders: Evidence from an online survey

*By Edgar Pacheco, PhD*[1]

To explore social media use in New Zealand, a sample of 1001 adults aged 18 and over were surveyed in November 2021. Participants were asked about the frequency of their use of different social media platforms (text message included). This report describes how often each of the nine social media sites and apps covered in the survey are used individually on a daily basis. Differences based on key demographics, i.e., age and gender, are tested for statistical significance, and findings summarised.

Findings show that Facebook, followed by SMS text, and Facebook Messenger are the tools most frequently used by adult New Zealanders. A significant association between age and frequent use of each of the tools explored in this study was also found. However, statistical association between gender and social media use was only found to be significant with some platforms.

## Points of interest

- Facebook remains as the most frequently social media platform used by adult New Zealanders. 67% use it daily. SMS text and Facebook Messenger are also commonly used.
- Younger adults tend to use social media more frequently than older adults. Instagram, Snapchat, and TikTok are especially common among those aged 18-29.
- The gap of social media use becomes bigger among older age groups. Those aged 50-64, and 65 and older are less likely to use these tools compared to their younger counterparts.
- In terms of gender, females were more likely to daily use Facebook, Facebook Messenger, and Instagram compared to males.
- Meanwhile, males use Twitter and LinkedIn more frequently than females.
- These findings may help to inform conversation and initiatives aiming at enhancing digital inclusion and digital transformation in New Zealand.

---

[1] Edgar Pacheco holds a PhD in Information Systems. Dr. Pacheco is currently an Adjunct Research Fellow at Victoria University of Wellington's School of Information Management, and a Senior Research Scientist (Social Sciences) at WSP New Zealand. His work has extensively covered aspects of digital technologies in relation to risks and opportunities, transition, and disability. Email: edgar.pacheco@vuw.ac.nz or e.pacheco1000@gmail.com. @edgarpachecob1.

## Findings

### *Overall frequency of social media use*

As Figure 1 shows, nearly 7 in 10 adult New Zealanders said they use Facebook once or more a day. When asked about SMS text, over half (55%) indicated they used this tool on a daily basis. A similar rate was found for users of Facebook Messenger (54%), while 32% said they use Instagram daily. Individual rates for the rest of the social media tools studied were lower.

***Figure 1. Social media use on a daily basis in 2019, 2020, and 2021***

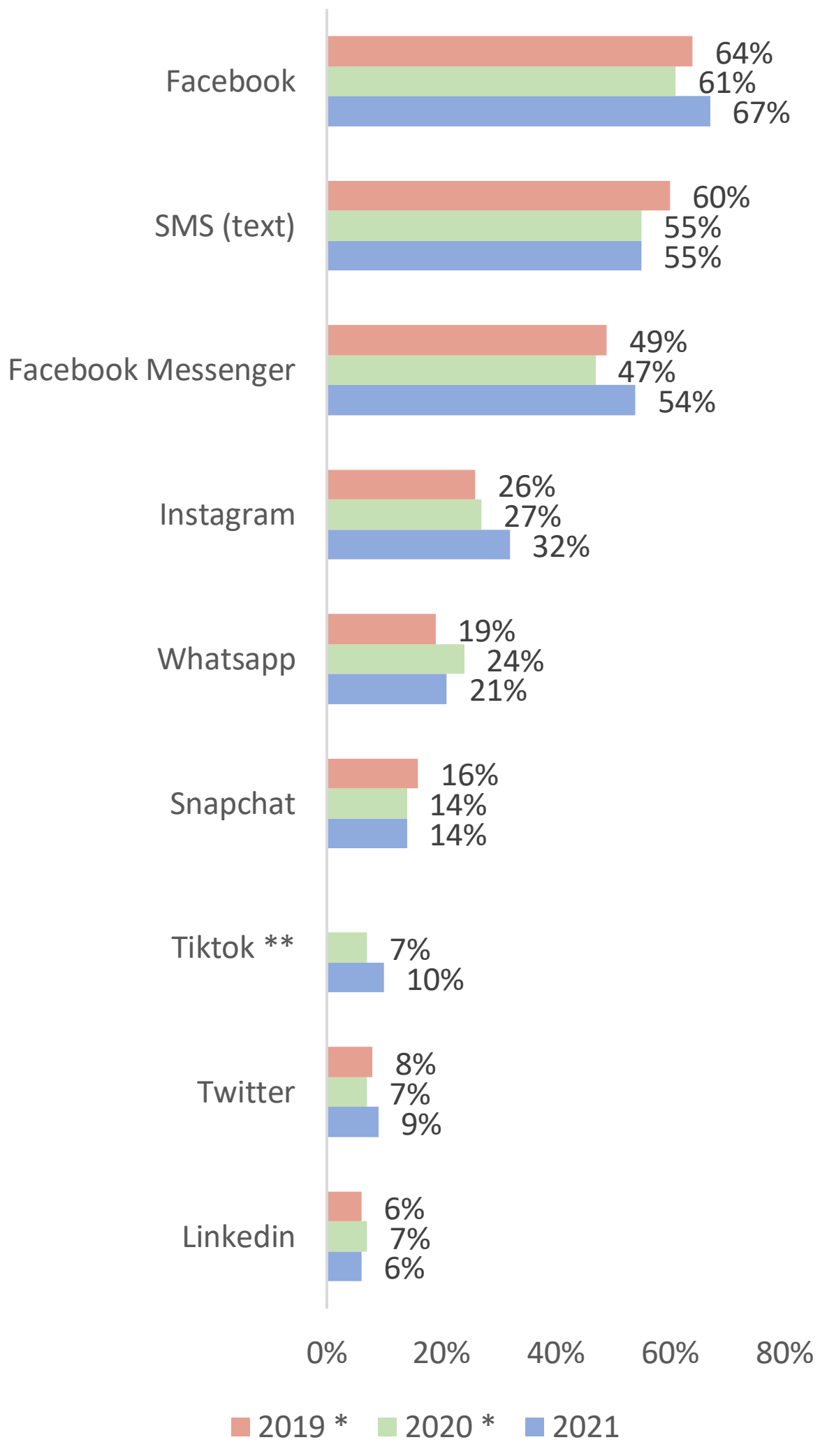

*Note: * for further details see InternetNZ (2019, 2020).*
*** not included in the 2019 survey.*

Figure 1 also compares findings with those from similar surveys conducted in 2019 and 2020 (see InternetNZ, 2019, 2020). Overall, there were no major changes in the trends for each social media platform. However, some small growth in the frequency of social media use was registered for Facebook, Instagram and Facebook Messenger, when compared with the 2020 findings.

### *Age differences in social media use*

The findings show that younger adults aged 18-29 use Instagram, Snapchat, and Tik Tok more frequently than their older counterparts. For instance, regarding Instagram, there was 16 percentual points difference between those aged 18-29 (59.7%) and those aged 30-49 (43.2%). The difference with those aged 65 and older was much bigger, 52 percentage points.

Also, the gap regarding daily use of Snapchat was 29 percentage points between participants aged 18-29 (39.8%) and those aged 30-39 (10.8%). Participants in the 18-29 age group were also more likely to frequently use Facebook Messenger, Twitter, and LinkedIn.

On the other hand, 30-49-year-old participants were more likely to frequently use Facebook, SMS text, and WhatsApp. 72.0% of participants in this group reported that they used Facebook once a day or more often. This was followed by those aged 18-29 (68.3%). Rates for participants aged 50-64, and 65 and older were lower but still significant, 59.7% and 60.8%, respectively.

Also, the data shows that SMS text is more commonly used by those aged 30-49 (64.6%), and 50-64 years old (60.9%).

Regarding WhatsApp, 27.6% of those aged 30-49 indicated they used this instant messaging app daily. For further details see Figure 2.

***Figure 2. Social media use on a daily basis in 2021 by age group***

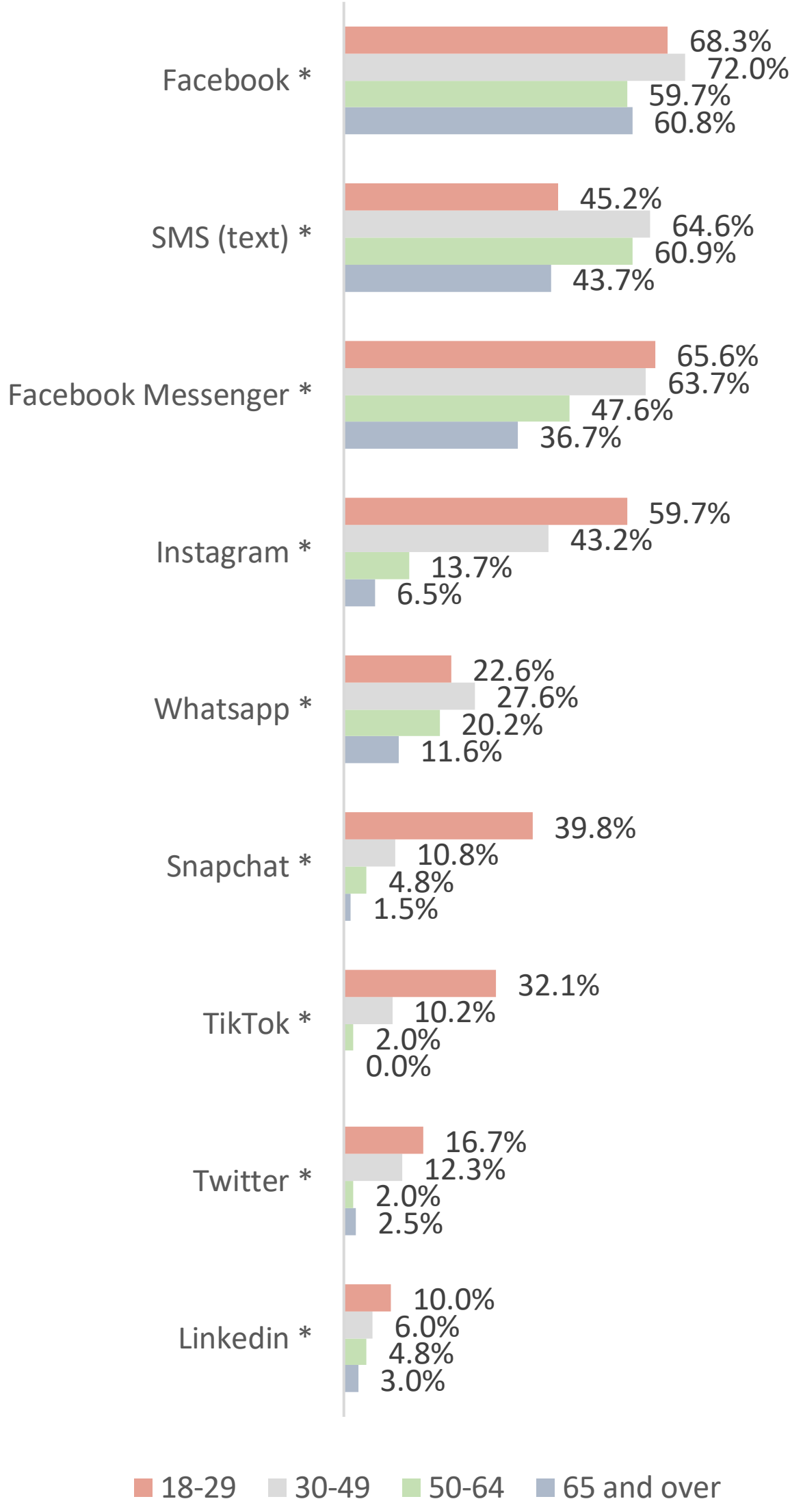

*Note: * p < .001*

### *Gender differences in social media use*

Statistically significant gender differences were found in five of the nine social media tools covered in this report. Females (74%), for instance, were more likely to use Facebook once a day or more often than males (58.0%). The difference was greater regarding frequent use of the Facebook Messenger app with 64.0% for females and 44.7% for males. Instagram was also more commonly used by females (38.9%) compared with males (25.2%).

Meanwhile, for males it was more common to use Twitter. 12.9% of male participants said they use this social media tool once a day or more, compared to 4.8% of females. LinkedIn was also more commonly used by males (7.5%) than females (4.6%). See Figure 3.

The study did not find a significant association between gender and the individual use of SMS text, Snapchat, WhatsApp, and Tik Tok ($p$ > .05).

***Figure 3. Social media use on a daily basis in 2021 by gender***

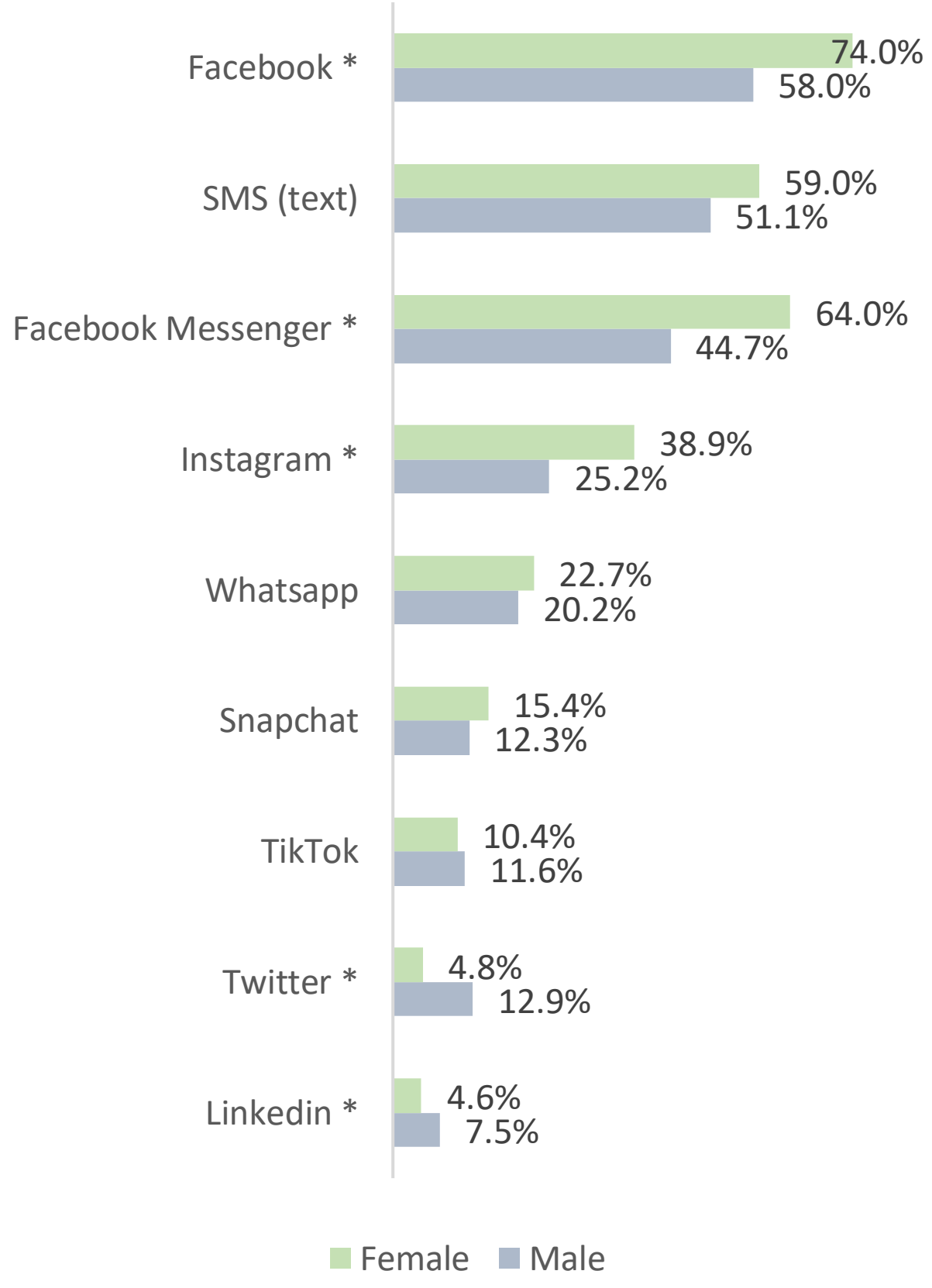

*Note: * p < .001*

## Concluding remarks

This short report presents insights about how often adult New Zealanders engage with different social media tools. Findings come from a sample of 1001 adult participants who took part in an online survey.

In terms of demographic differences, and similar to recent international research (see Auxier & Anderson, 2021), a age gap has been identified. Compared with older age groups, younger people in our sample are significantly more likely to daily use social media tools, in particular Instagram, Snapchat, and Tik Tok.

Research has pointed out that older adults in New Zealand are less likely to have access to the internet (Grimes & White, 2019; MSD, 2016), and are more likely to never use it (Diaz Andrade et al., 2021; MBIE & DIA, 2017). The findings in this report complements this evidence as it shows that the age gap also involves social media use among those already connected older New Zealanders. This has implications for current efforts supporting digital inclusion, understood as equitable opportunities to participate in society via digital technologies (digital.govt.nz, 2020). As more government agencies and private businesses rely on digital tools to improve their processes and services, older internet users may be left behind by the opportunities of these innovation and changes.

While research (see Pacheco & Melhuish, 2018, 2019; Vogels et al., 2022) suggests that teenagers aged 17 or under are significantly less likely to use Facebook, this report shows that this is not the case among adults. Facebook is the most used platform by adult New Zealanders, including those in the 18-29 age group (nearly 7 in 10).

Another interesting finding concerns to gender. Clearly, the findings show that males and females engage differently with some specific social media sites and apps. The former, males, are more likely to use Twitter and LinkedIn while the latter, females, engage more often with tools such as Facebook, Facebook Messenger, and Instagram. This pattern of social media preferences in terms of gender has already be seen among children and teenagers in New Zealand (Pacheco & Melhuish, 2018, 2019).

This report is only a snapshot of social media use in New Zealand. Thus, more longitudinal research is needed to understand trends in this regard. Despite this, we believe the findings add to current policy and research attempts to better understand New Zealander's complex and varied engagement with digital technologies.

## Methodology

We conducted secondary data analysis (Sturgis et al., 2009), of the *New Zealand's Internet Insights*, a survey conducted annually by InternetNZ. The survey was administered online to 1,001 New Zealanders aged 18 and over by Kantar Public. The market research company used a combination of pre-survey quotas and post survey weighting to ensure results are representative of all New Zealanders by key demographics such as age and gender. Online surveys are not only cost-effective and easier to administer but also their use as a technique for data collection is growing in the social sciences (Sue & Ritter, 2012). In New Zealand recent exploratory research using online surveys has provided relevant evidence about diverse aspects related to the opportunities and risks of digital technologies (Pacheco & Melhuish, 2017, 2019, 2020, 2021).

Consent to take part in the study was obtained from all participants, who also had the right to withdraw from it at any time.

Fieldwork was conducted between the 3 and 17 November 2021. The maximum margin of error on the total sample is +/- 3.1% at the 95% confidence interval.

For demographic distribution of the sample see Table 1 below.

***Table 1. Sample demographic distribution (N=1001).***

| Levels | Counts | % |
|---|---|---|
| **Age group** | | |
| 18-29 | 221 | 22.1 |
| 30-49 | 333 | 33.3 |
| 50-64 | 248 | 24.8 |
| 65 and over | 199 | 19.9 |
| **Gender** | | |
| Male | 481 | 48.1 |
| Female | 519 | 51.8 |
| Gender diverse * | 1 | 0.1 |

*Note: * excluded from analysis due to small number.*

Data are presented as percentages. All analyses were performed using the Jamovi software, version 2.3 (The jamovi project, 2022). To test significance of association between categorical variables we used Chi-square of Independence. A *p* value equal or less than 0.05 was used to indicate statistical significance.

## Acknowledgments

This report would not have been possible without the support of InternetNZ. InternetNZ is a not-for-profit open membership organisation and the designated manager for the .nz top level internet domain. It also supports the development of New Zealand's internet through policy, community grants, research, and events.